\documentclass[10pt,english,british,tightenlines,eqsecnum,
floats,aps,amsmath,amssymb,nofootinbib,superscriptaddress,prd,showpacs,showkeys]
              {revtex4}

\usepackage[latin9]{inputenc}
\usepackage{color}
\usepackage{babel}
\usepackage{amsmath}
\usepackage{amssymb}
\usepackage{graphicx}
\usepackage{esint}
\usepackage[unicode=true,pdfusetitle,
 bookmarks=true,bookmarksnumbered=false,bookmarksopen=false,
 breaklinks=false,pdfborder={0 0 1},backref=false,colorlinks=false]
 {hyperref}
\makeatletter
\@ifundefined{textcolor}{}
{%
 \definecolor{BLACK}{gray}{0}
 \definecolor{WHITE}{gray}{1}
 \definecolor{RED}{rgb}{1,0,0}
 \definecolor{GREEN}{rgb}{0,1,0}
 \definecolor{BLUE}{rgb}{0,0,1}
 \definecolor{CYAN}{cmyk}{1,0,0,0}
 \definecolor{MAGENTA}{cmyk}{0,1,0,0}
 \definecolor{YELLOW}{cmyk}{0,0,1,0}
}

\@ifundefined{date}{}{\date{16 Nov 2023}}
\usepackage{babel}

\usepackage{euscript}\usepackage{epsfig}

\usepackage{amsfonts}

\usepackage{fancyhdr}

\makeatother

\usepackage{amsthm}
\theoremstyle{plain}
\newtheorem*{remark*}{Remark}
\usepackage{cleveref}

\begin{document}

\title{Noncompactified Kaluza--Klein theories and Anisotropic Kantowski-Sachs Universe}

\author{S. M. M. Rasouli}

\email{mrasouli@ubi.pt}

\affiliation{Departamento de F\'{i}sica,
Centro de Matem\'{a}tica e Aplica\c{c}\~{o}es (CMA-UBI),
Universidade da Beira Interior,
Rua Marqu\^{e}s d'Avila
e Bolama, 6200-001 Covilh\~{a}, Portugal}

\affiliation{Department of Physics, Qazvin Branch, Islamic Azad University,
Qazvin 341851416, Iran}

\begin{abstract}
We provide an overview of noncompactified Kaluza--Klein theories.
The space--time-matter theory (or induced-matter theory) and the
modified Brans--Dicke theory are discussed.
Finally, an extended version of the Kantowski--Sachs anisotropic
model is investigated as a cosmological application of the latter.
\end{abstract}

\medskip


\maketitle

\section{Introduction}
\label{introduction}
\indent
An outline of the original Kaluza--Klein (KK) theory is given in this section. Subsequently, in a few succinct sentences, we concentrate on the noncompactified KK theories, their motives, and how they differ from their compactified counterparts. For a more in-depth analysis, see \cite{OW97}.

The close relationship between Minkowski's four-dimensional
(4D) spacetime and Maxwell's unification of electricity and
magnetism may have been the main inspiration for N\"{o}rdstrom
and Kaluza, who were the first physicists to attempt to unify gravity and electromagnetism using a 5D framework.
Both have assumed that all derivatives with respect to
the fifth coordinate $x^4\equiv l$ must be zero (cylinder condition) \cite{OW97}.

Let us give a brief overview of a generalized version of the
Kaluza mechanism, for a detailed study see \cite{OW97, Williams15}.
(i) Einstein's field equations were assumed in five
dimensions along with the Kaluza's
first key assumption: ``\textit{the universe in higher dimensions is empty.}'' \cite{OW97}.
(ii) The definitions of the Christoffel symbols and the 5D Ricci
tensor are taken exactly as they are defined in four dimensions (\textit{Kaluza's
second key assumption}). (iii) The cylinder condition has been assumed as the \textit{third key assumption}.
By combining the aforementioned assumptions with the specific 5D line--element ${\cal G}_{ab}^{^{\rm KK}}$
(Throughout this review paper, we take the signature of the 4D metric as (- + + +).
Moreover, the small Latin indices and the Greek indices run over
$0,1,2,3,4$ and $0,1,2,3$, respectively.) with components
\begin{equation}\label{KK-metric}
{\cal G}_{\mu\nu}^{^{\rm KK}}=g_{\mu\nu}+\psi^2 A_\mu A_\nu,\hspace{5mm}
{\cal G}_{5\nu}^{^{\rm KK}}={\cal G}_{\nu 5}^{^{\rm KK}}=\psi^2 A_\nu, \hspace{5mm} {\cal G}_{55}^{^{\rm KK}}=\psi^2,
\end{equation}
(Where $g_{\mu\nu}$, $A_\mu$ and $\psi$ denote the 4D metric tensor, electromagnetic
potential and a scalar field, respectively.) 15 field equations
are obtained on a 4D hypersurface. With KK assumption, $\psi=1$, one
obtains the Einstein equations (with the electromagnetic
energy-momentum tensor) and Maxwell equations on the hypersurface.

The fifth dimension, as mentioned in Kaluza's mechanism, exists, but according to
Kaluza's first key assumption, the derivatives of physical quantities
with respect to the fifth coordinate $l$ should vanish.
However, in the \textit{compactified KK theory}, the fifth coordinate
was assumed to be a lengthlike with two properties:
a circular topology and a small scale, in accordance with \textit{Klein's assumption}.
Given the former, any quantity can become periodic, so the only modes
that are independent of the fifth coordinate are observables, as required by \cite{OW97}.

In the noncompactified KK theories, which are the main
focus of this review, the idea that the new
coordinates are physical still remains, but the
compactified approach is generalized by omitting the Kaluza's
third key assumption. This means that all the
physical quantities can depend on the extra coordinate.
Therefore, in these noncompactified frameworks, the
equations of motion are modified by the dependence on extra
coordinate. Consequently, not only the electromagnetic
radiation emerge from geometry, but also a very general kind of matter.
It is worth noting that in  noncompactified theories the extra
coordinates should not necessarily be assumed to be lengthlike.

One of the most well-known noncompactified KK theories is
the space-time-matter theory or the induced-matter
theory (IMT) \cite{PW92,wesson1999space}, which is briefly introduced in the next section.
In the IMT, using Einstein's general theory of relativity
(GR) as the underlying theory and using an appropriate
reduction method, it has been shown that 5D KK equations without
sources generates Einstein equations with the induced energy-momentum tensor (EMT).
Cosmological/astrophysical applications of the IMT (or its extended versions) are
widespread in the literature, see~\cite{wesson1999space,OW97,LMPJVM21}, and references therein.
For instance, interesting solutions were
obtained in Ref.~\cite{doroud2009class} by considering a conformally flat bulk space. Then, the energy conditions
associated with this model have also been studied in \cite{Rasouli:2010zz}.

Recently, generalized versions of the IMT have been established by
applying some alternative theories to GR as the underlying theory:
the modified S\'aez--Ballester theory (MSBT) \cite{RM18,RPSM20,Rasouli:2022tjn} and
the Modified Brans--Dicke Theory (MBDT) \cite{RFM14,Amani:2022arl} are two important examples.
One of the main goals of this review will be the latter and its cosmological application.

This review paper is structured as follows. In the next section we give a
brief overview of the IMT. In the Section \ref{MBDT} we describe the
 field equations of the MBDT. In the \ref{Bulk} section, considering
 the $5D$ empty Kantowski-Sachs (KS) universe, the BD field equations
 (we call it BD-KS cosmology) and a new time coordinate, we get the
 corresponding exact solutions. Then the corresponding constraints
 among the parameters of the model and some special cases are presented.
 The Section \ref{red-cosmology} discusses the properties of
 physical quantities associated with the BD-KS cosmology of the
 4D hypersurface. Finally, in Section \ref{Conclusions} we present a summary and further discussion.

\section{Five-dimensional Ricci--Flat Space and the IMT}
\label{IMT}
\indent

In this section, we present a brief overview of the IMT, for a detailed
study see~\cite{wesson1992kaluza} and the references therein.

Considering GR as the underlying theory in an empty 5D spacetime, and applying a
reduction method, the~field equations of GR with sources
can be set up on a 4D hypersurface.

By considering the 5D metric as
\begin{equation}\label{global-metric-1}
dS^{2}={\cal G}_{ab}(x^c)dx^{a}dx^{b},
\end{equation}
where our 4D universe can be embedded locally and isometrically,
is \mbox{taken as}
\begin{equation}\label{global-metric}
dS^{2}=g_{\mu\nu}(x^\alpha,l)dx^{\mu}dx^{\nu}+
\epsilon\psi^2\left(x^\alpha,l\right)dl^{2}.
\end{equation}
In equation \eqref{global-metric},
$\psi=\psi\left(x^\alpha,l\right)$ is a scalar field and
$\epsilon=\pm1$ (where $\epsilon^2=1$) is an indicator that allows the
extra dimension to be selected as time--like or space--like.\rlap\footnote{In order to keep in touch with the content of the
original papers in each section, we use the same units
that were included therein. For example, in~this section we
take the same units used in~\cite{wesson1992kaluza}.}
The metric \eqref{global-metric} is the KK metric \eqref{KK-metric} with $A_\mu=0$
(further justification was presented in \cite{OW97}).

The $\alpha\beta-$, $\alpha4-$ and $44-$parts of the Ricci tensor $R^{^{(5)}}_{ab}$ are:
\begin{eqnarray}\label{ricci-tensor-5,4}
R^{^{(5)}}_{\alpha\beta}\!\!\!&=&\!\!\!
R^{^{(4)}}_{\alpha\beta}-\frac{{\cal D}_\alpha{\cal
D}_\beta\psi}{\psi}
+\frac{\epsilon}{2\psi^2}\left(\frac{{\overset{*}\psi}{\overset{*}g}_{\alpha\beta}}{\psi}
-{\overset{**}g}_{\alpha\beta}+g^{\lambda\mu}{\overset{*}g}_{\alpha\lambda}{\overset{*}g}_{\beta\mu}
-\frac{1}{2}g^{\mu\nu}{\overset{*}g}_{\mu\nu}{\overset{*}g}_{\alpha\beta}\right),
\\\nonumber
\\
\label{R_4-alpha}
R^{^{(5)}}_{4\alpha}\!\!\!&=&\!\!\!\psi{\cal
D}_\beta P^{\beta}{}_{\alpha},
\\\nonumber
\\
\label{R_44}
R^{^{(5)}}_{_{44}}\!\!\!&=&\!\!\!-\epsilon\psi{\cal
D}^2\psi-\frac{1}{4}
{\overset{*}g}^{\lambda\beta}{\overset{*}g}_{\lambda\beta}
-\frac{1}{2}g^{\lambda\beta}{\overset{**}g}_{\lambda\beta}
+\frac{\overset{*}\psi}{2\psi}g^{\lambda\beta}{\overset{*}g}_{\lambda\beta},
\end{eqnarray}
where
\begin{equation}\label{P-mono}
P_{\alpha\beta}\equiv\frac{1}{2
\psi}\left({\overset{*}g}_{\alpha\beta}
-g_{\alpha\beta}g^{\mu\nu}{\overset{*}g}_{\mu\nu}\right),
\end{equation}
 $\overset{*}A\equiv \frac{\partial A}{\partial l}$, ${\cal D}_\alpha$ denotes the covariant derivative on the
4D hypersurface and ${\cal D}^2\equiv{\cal D}^\alpha{\cal D}_\alpha$.

 Assuming a 5D space-time without sources~\cite{wesson1992kaluza,OW97,wesson1999space}
 (namely, $G^{^{(5)}}_{ab}=0=R^{^{(5)}}_{ab}$, where $R^{^{(5)}}_{ab}$ is the Ricci tensor) and
 defining a 4D hypersurface $\Sigma_4$,
 where $l=l_0={\rm constant}$, and
 \begin{eqnarray}\label{induce-met}
g_{\mu\nu}(x^\alpha)= {\cal G}^{^{(4)}}(x^\alpha,l_0),
 \end{eqnarray}
equations \eqref{ricci-tensor-5,4}--\eqref{R_44} yield
\begin{eqnarray}\label{2-ricci-tensor-5,4}
R^{^{(4)}}_{\alpha\beta}\!\!\!&=&\!\!\!
\frac{{\cal D}_\alpha{\cal
D}_\beta\psi}{\psi}
-\frac{\epsilon}{2\psi^2}\left[\frac{{\overset{*}\psi}{\overset{*}g}_{\alpha\beta}}{\psi}
-{\overset{**}g}_{\alpha\beta}+g^{\lambda\mu}{\overset{*}g}_{\alpha\lambda}{\overset{*}g}_{\beta\mu}
-\frac{1}{2}g^{\mu\nu}{\overset{*}g}_{\mu\nu}{\overset{*}g}_{\alpha\beta}\right],
\\\nonumber
\\
\label{2-R_4-alpha}
{\cal D}_\beta P^{\beta}{}_{\alpha}\!\!\!&=&0,
\\\nonumber
\\
\label{2-R_44}
\epsilon\psi{\cal D}^2\psi\!\!\!&=&\!\!\!-\frac{1}{4}
{\overset{*}g}^{\lambda\beta}{\overset{*}g}_{\lambda\beta}
-\frac{1}{2}g^{\lambda\beta}{\overset{**}g}_{\lambda\beta}
+\frac{\overset{*}\psi}{2\psi}g^{\lambda\beta}{\overset{*}g}_{\lambda\beta}.
\end{eqnarray}

Equations \eqref{2-ricci-tensor-5,4}--\eqref{2-R_44} ``\textit{ form the basis of five-dimensional noncompactified KK theory}'' \cite{OW97}.

 Employing \eqref{2-ricci-tensor-5,4}, \eqref{2-R_44}, and~  $(\delta^\mu_\nu)_{,4}=0=g^{\mu\beta}g^{\sigma\lambda}{\overset{*}g}_{\lambda\beta}{\overset{*}g}^{\mu\sigma}
  +{\overset{*}g}_{\mu\sigma}{\overset{*}g}_{\mu\sigma}$, we get $R^{^{(4)}}=g^{\alpha\beta}R^{^{(4)}}_{\alpha\beta}$ as
\begin{equation}\label{4-Ricci}
R^{^{(4)}}=\frac{\epsilon}{4\psi^2}\left[{\overset{*}g}^{\mu\nu}{\overset{*}g}_{\mu\nu}+\left(g^{\mu\nu}{\overset{*}g}_{\mu\nu}\right)^2\right].
\end{equation}

If we define the EMT in four dimensions as $T_{\alpha\beta}^{^{[\rm IMT]}}\equiv R^{^{(4)}}_{\alpha\beta}-1/2 R^{^{(4)}}g_{\alpha\beta}$, then from \eqref{2-ricci-tensor-5,4} and \eqref{4-Ricci}, we can easily get\vspace{-3pt}
 \begin{multline}\label{IMTmatt.def}
T_{\alpha\beta}^{^{[\rm IMT]}}\equiv
\frac{{\cal D}_\alpha{\cal D}_\beta\psi}{\psi}
-\frac{\epsilon}{2\psi^{2}}\left(\frac{{\overset{*}\psi}{\overset{*}g}_{\alpha\beta}}{\psi}-{\overset{**}g}_{\alpha\beta}
+g^{\lambda\mu}\overset{*}{g}_{\alpha\lambda}{\overset{*}g}_{\beta\mu}
-\frac{1}{2}g^{\mu\nu}\overset{*}{g}_{\mu\nu}{\overset{*}g}_{\alpha\beta}\right)\\-\frac{\epsilon g_{\alpha\beta}}{8\psi^2}
\left[{\overset{*}g}^{\mu\nu}{\overset{*}g}_{\mu\nu}
+\left(g^{\mu\nu}{\overset{*}g}_{\mu\nu}\right)^{2}\right].
 \end{multline}

We can consider $T_{\alpha\beta}^{^{[\rm IMT]}}$ as an EMT of our 4D
universe, known as {\it induced-matter} in the noncompactified KK theory.
Let us be more precise. This induced matter is actually a manifestation of the pure geometry of the higher-dimensional space-time.
It is worth noting that the equations on a 4D
hypersurface, i.e., $G_{\alpha\beta}^{^{(4)}}=T_{\alpha\beta}^{^{[\rm IMT]}}$, are automatically
included in the corresponding 5D vacuum counterparts $G^{^{(5)}}_{ab}=0$.

\section{Modified Brans-Dicke~Theory}
\label{MBDT}

In this section we would like to give an overview of the framework established in Ref.\cite{RFM14}.
Here, we also consider the metric \eqref{global-metric} and the
noncompact extra dimension.
Moreover, equations \eqref{ricci-tensor-5,4}--\eqref{P-mono} are valid for our herein framework.
It should be noted, however, that since the BD scalar field can
play the role of a higher--dimensional matter, equations
$G^{^{(5)}}_{ab} =0=R^{^{(5)}}_{ab}=0$ are generally no longer valid.

The action of the Brans--Dicke (BD) theory in five dimensions in the Jordan
frame, in analogy with its standard counterpart, can be written as
\begin{equation}\label{(D+1)-action}
{\cal S}^{^{(5)}}_{\rm BD}=\int d^{^{5}}x \sqrt{\Bigl|{}{\cal G}^{(5)}}\Bigl| \,\left[\phi
R^{^{(5)}}-\frac{\omega}{\phi}\, {\cal G}^{ab}\,(\nabla_a\phi)(\nabla_b\phi)+16\pi\,
L\!^{^{(5)}}_{_{\rm matt}}\right],
\end{equation}
where $\phi$ is the BD scalar field, $\omega$ is a
dimensionless parameter
and $\nabla_a$ denotes the covariant derivative in the $5D$ space-time, respectively.

The BD field equations derived from the action (\ref{(D+1)-action}) are:
\begin{equation}\label{(D+1)-equation-1}
G^{^{(5)}}_{ab}=\frac{8\pi}{\phi}\,T^{^{(5)}}_{ab}+\frac{\omega}{\phi^{2}}
\left[(\nabla_a\phi)(\nabla_b\phi)-\frac{1}{2}{\cal G}_{ab}(\nabla^c\phi)(\nabla_c\phi)\right]
+\frac{1}{\phi}\Big(\nabla_a\nabla_b\phi-{\cal G}_{ab}\nabla^2\phi\Big),
\end{equation}
\begin{equation}\label{(D+1)-equation-2}
\frac{2\omega}{\phi}\nabla^2\phi
-\frac{\omega}{\phi^{^{2}}}{\cal G}^{ab}(\nabla_a\phi)(\nabla_b\phi)+R^{^{(5)}}=0,
\end{equation}
where ~$\nabla^2\equiv\nabla_a\nabla^a$.
On the other hand, from equation~(\ref{(D+1)-equation-1}), we obtain
\begin{equation}\label{(D+1)-equation-3}
R^{^{(5)}}=-\frac{16\pi\,T^{^{(5)}}}{3\phi}+\frac{\omega}{\phi^{2}}(\nabla^c\phi)(\nabla_c\varphi)
+\frac{8}{3}\left(\frac{\nabla^2\phi}{\phi}\right).
\end{equation}
where $T^{^{(5)}}={\cal G}^{ab}T^{^{(5)}}_{ab}$.

Substituting $R^{^{(5)}}$ from equation (\ref{(D+1)-equation-3}) into (\ref{(D+1)-equation-2}), we obtain
\begin{equation}\label{(D+1)-equation-4}
\nabla^2\phi=\frac{8\pi T^{^{(5)}}}{3\omega+4}.
\end{equation}

A suitable reduction method was used to establish the MBDT~\cite{RFM14},
see also \cite{Rasouli:2021xqz} for a corrected version of some field equations in the $D$ dimensions.
Therefore, let us confine ourselves in this paper
to presenting only a summary of a special case where
there is no higher--dimensional ordinary matter, i.e., $L\!^{^{(5)}}_{_{\rm matt}}=0$.
Concretely, considering \eqref{global-metric} as a background metric, equations~\eqref{(D+1)-equation-1} and
(\ref{(D+1)-equation-4}) yield four sets of effective field equations on the hypersurface:
\begin{enumerate}
  \item An equation associated with the scalar field $\psi$:
\begin{eqnarray}\label{D2say-BD}
\frac{{\cal D}^2\psi}{\psi}=-\frac{({\cal D}_\alpha\psi)({\cal D}^\alpha\phi)}{\psi\phi}
-\frac{\epsilon}{2\psi^2}
\left(g^{\lambda\beta}{\overset{**}g}_{\lambda\beta}+\frac{1}{2}{\overset{*}g^{\lambda\beta}}
{\overset{*}g}_{\lambda\beta}-\frac{g^{\lambda\beta}{\overset{*}g}_{\lambda\beta}{\overset{*}\psi}}{\psi}\right)
-\frac{\epsilon}{\psi^2\phi}
\left[\overset{**}\phi+\overset{*}\phi\left(\frac{\omega\overset{*}\phi}{\phi}
-\frac{\overset{*}\psi}{\psi}\right)\right].
\end{eqnarray}
  \item
  An effective field equation that is the counterpart of \eqref{(D+1)-equation-1}:
    \begin{multline}\label{BD-DD}
G_{\mu\nu}^{^{(4)}}=\frac{8\pi T^{^{\rm [MBDT]}}_{\mu\nu}}{\phi}+
\frac{\omega}{\phi^2}\left[\left({\cal D}_\mu\phi\right)\left({\cal D}_\nu\phi\right)-
\frac{1}{2}g_{\mu\nu}({\cal D}_\alpha\phi)({\cal D}^\alpha\phi)\right]\\
+\frac{1}{\phi}\left({\cal D}_\mu{\cal D}_\nu\phi- g_{\mu\nu}{\cal D}^2\phi\right)
-g_{\mu\nu}\frac{V\left(\phi\right)}{2\phi}.
\end{multline}
 where we obtain the~induced potential $V(\phi)$ from
 equation \eqref{v-def} (see below); the effective energy--momentum
tensor $ T^{^{\rm [MBDT]}}_{\mu\nu}$, in turn, consists of three parts:
\begin{eqnarray}\label{matt.def}
\frac{8\pi}{\phi} T^{^{\rm [MBDT]}}_{\mu\nu}\equiv T_{\mu\nu}^{^{[\rm IMT]}}
+\frac{1}{\phi}T_{\mu\nu}^{^{[\phi]}}+\frac{V(\phi)}{2\phi}g_{\mu\nu}.
\end{eqnarray}
Regarding equation \eqref{matt.def}, it should be noted that the
term $T_{\mu\nu}^{^{[\rm IMT]}}$ is the same effective matter obtained
in the IMT for which GR was considered as an underlying theory.
However, for the BD theory, due to the presence of the BD scalar field, an
extra effective energy momentum tensor is also induced on the hypersurface:
\begin{eqnarray}\label{T-phi}
T_{\mu\nu}^{^{[\phi]}}\equiv
\frac{\epsilon\overset{*}\phi}{2\psi^2}\left[\overset{*}g_{\mu\nu}
+g_{\mu\nu}\left(\frac{\omega\overset{*}\phi}{\phi}-g^{\alpha\beta}
{\overset{*}g}_{\alpha\beta}\right)\right].
\end{eqnarray}
It is seen that $T_{\mu\nu}^{^{[\phi]}}$ depends on the first derivatives of $\phi$ with respect to the fifth coordinate.

\item
The wave equation associated with the MBDT:
\begin{eqnarray}\label{D2-phi}
{\cal D}^2\phi=\frac{1}{2\omega+3}
\left[8\pi T^{^{\rm [MBDT]}}+\phi\,\frac{dV\left(\phi\right)}{d\phi}
-2V\left(\phi\right)\right],
\end{eqnarray}
where $V(\phi)$ is obtained from
\begin{multline}\label{v-def}
\phi \frac{dV(\phi)}{d\phi}\equiv -2(\omega+1)
\left[\frac{({\cal D}_\alpha\psi)({\cal D}^\alpha\phi)}{\psi}
+\frac{\epsilon}{\psi^2}\left(\overset{**}\phi-
\frac{\overset{*}\psi\overset{*}\phi}{\psi}\right)\right]\\-\frac{\epsilon\omega\overset{*}\phi}{\psi^2}
\left(\frac{\overset{*}\phi}{\phi}+g^{\mu\nu}\overset{*}g_{\mu\nu}\right)
+\frac{\epsilon\phi}{4\psi^2}
\Big[\overset{*}g^{\alpha\beta}\overset{*}g_{\alpha\beta}
+(g^{\alpha\beta}\overset{*}g_{\alpha\beta})^2\Big].
\end{multline}
\item
 An effective equation that is counterpart to the conservation equation introduced in the IMT:
\begin{equation}\label{P-Dynamic}
G^{^{(5)}}_{\alpha 4}= \psi{\cal D}_\beta P^{\beta}{}_{\alpha}=
\frac{\omega\overset{*}\phi({\cal D}_\alpha\phi)}{\phi^2}
+\frac{{\cal D}_\alpha\overset{*}\phi}{\phi}
-\frac{\overset{*}g_{\alpha\lambda}\left({\cal D}^\lambda\phi\right)}{2 \phi}-
\frac{\overset{*}\phi({\cal D}_\alpha\psi)}{\phi\psi}.
\end{equation}

\end{enumerate}

It is worthy noting that equations~(\ref{BD-DD}) and (\ref{D2-phi}) can be retrieved
 from the action
\begin{equation}
{\cal S}^{^{(5)}}_{\rm BD}=\int d^{^{\,4}}\!x \sqrt{-g}\,\left[\phi
R^{^{(4)}}-\frac{\omega}{\phi}\, g^{\alpha\beta}\,({\cal
D}_\alpha\phi)({\cal D}_\beta\phi)-V(\phi)+16\pi\,
L\!^{^{(4)}}_{_{\rm matt}}\right],
\end{equation}
 where
$\sqrt{-g}T^{^{\rm [MBDT]}}_{\mu\nu}\equiv 2\delta\left( \int d^{^{\,4}}\!x
\sqrt{-g}\,\,L\!^{^{(4)}}_{_{\rm matt}}\right)/\delta g^{\alpha\beta}$.

In summary, using (\ref{global-metric}) as the background line element
applying a suitable reduction method, the effective MBDT field equations, i.e.,
~(\ref{D2say-BD}), (\ref{BD-DD}), (\ref{D2-phi}) and~(\ref{P-Dynamic}), are
 obtained from equations~(\ref{(D+1)-equation-1}) and (\ref{(D+1)-equation-4}).
It is worth noting that both $T^{^{\rm [MBDT]}}_{\mu\nu}$ and $V(\phi)$ can be
considered as fundamental quantities rather than ad~hoc
phenomenological~assumptions taken in the conventional generalized BD frameworks.

\section{Exact BD-KS anisotropic vacuum solutions in five-dimensions}
\label{Bulk}
In this section, let us review the solutions of the $5D$ BD field equations
\eqref{(D+1)-equation-1} and \eqref{(D+1)-equation-4} for the
extended anisotropic Kantowski-Sachs metric~\cite{Rasouli:2018owa}, i.e.,
\begin{eqnarray}\label{DO-metric}
dS^{2}=-dt^{2}+a^{2}(t)dr^2+
b^2(t)\left(d\theta^2+ \sin^2\theta d\phi^2\right)+\epsilon \psi^2(t)dl^{2},
\end{eqnarray}
where $t$ is the cosmic time;
$a(t)$, $b(t)$ and $\psi(t)$ are cosmological scale factors.

Assuming that the BD scalar field depends only on time, i.e., $\phi=\phi(t)$, and
 using equations~(\ref{(D+1)-equation-1}), (\ref{(D+1)-equation-4}) and (\ref{DO-metric}), we obtain
\begin{eqnarray}\label{dot-1}
\frac{\ddot{\phi}}{\phi}\!\!&+&\!\!\frac{\dot{\phi}}{\phi}\left(\frac{\dot{a}}{a}+\frac{2\dot{b}}{b}+\frac{\dot{\psi}}{\psi}\right)=0,\\\nonumber\\
\label{dot-2}
\frac{\dot{b}}{b}\left(\frac{2\dot{a}}{a}+\frac{\dot{b}}{b}\right)\!\!&+&\!\!\frac{\dot{\phi}}{\phi}
\left[\frac{\dot{a}}{a}+\frac{2\dot{b}}{b}-\frac{\omega}{2}\left(\frac{\dot{\phi}}{\phi}\right)+\frac{\dot{\psi}}{\psi}\right]
+\frac{\dot{\psi}}{\psi}\left(\frac{\dot{a}}{a}+\frac{2\dot{b}}{b}\right)+\frac{1}{b^2}=0,\\\nonumber\\
\label{dot-3}
\frac{\ddot{b}}{b}\!\!&+&\!\!\frac{\dot{b}}{b}\left(\frac{\dot{a}}{a}+\frac{\dot{b}}{b}+\frac{\dot{\phi}}{\phi}\right)
+\frac{1}{2}\left[\frac{\ddot{\psi}}{\psi}+\frac{\dot{\psi}}{\psi}\left(\frac{\dot{a}}{a}+
\frac{4\dot{b}}{b}+\frac{\dot{\phi}}{\phi}\right)\right]+\frac{1}{b^2}=0,\\\nonumber\\
\label{dot-4}
\frac{\ddot{a}}{a}\!\!&+&\!\!\frac{\dot{a}}{a}\left(\frac{2\dot{b}}{b}+\frac{\dot{\phi}}{\phi}\right)
+\frac{1}{2}\left[\frac{\ddot{\psi}}{\psi}+\frac{\dot{\psi}}{\psi}\left(\frac{3\dot{a}}{a}+
\frac{2\dot{b}}{b}+\frac{\dot{\phi}}{\phi}\right)\right]=0,\\\nonumber\\
\label{dot-5}
\frac{\ddot{a}}{a}\!\!&+&\!\!\frac{2\ddot{b}}{b}+\frac{\dot{b}}{b}\left(\frac{2\dot{a}}{a}+\frac{\dot{b}}{b}\right)
+\frac{\dot{\phi}}{\phi}\left[\frac{\omega}{2}\left(\frac{\dot{\phi}}{\phi}\right)-\frac{\dot{\psi}}{\psi}\right]+\frac{1}{b^2}=0,
\end{eqnarray}
where an overdot denotes a derivative
with respect to $t$.

Among~(\ref{dot-1})-(\ref{dot-5}), only four equations are independent, where we also have four
unknowns $a$, $b$, $\phi$ and $\psi$.
To solve these equations we introduce a new time coordinate $\eta$ related to $t$ as
\begin{eqnarray}
\label{conformal}
dt=bd\eta.
\end{eqnarray}

After some manipulation, it is easy to
show that equations~(\ref{dot-1})-(\ref{dot-5}) can be rewritten as
\begin{eqnarray}
\label{prime-1}
ab\psi\phi'&=&c_1,\\\nonumber
\\
\label{prime-2}
Z''+ Z\!&=&\!0,\hspace{10mm} {\rm where}\hspace{10mm}Z\equiv ab\phi\psi,\\\nonumber
\\
\label{prime-3}
(XZ)'+2 Z\!&=&\!0,\hspace{10mm} {\rm where}\hspace{10mm} X\equiv [{\rm ln}(ab^2)]',\\\nonumber
\\
\label{prime-4}
[{\rm ln}(YZ)]'\!&=&\!0,\hspace{10mm} {\rm where}\hspace{10mm}Y\equiv[{\rm ln}(a\psi^{\frac{1}{2}})]',\\\nonumber\\
\label{prime-5}
\frac{b'}{b}\left(\frac{2a'}{a}+\frac{b'}{b}\right)\!&+&\!\frac{\phi'}{\phi}
\left[\frac{\psi'}{\psi}-\frac{\omega}{2}\left(\frac{\phi'}{\phi}\right)\right]
+\left(\frac{a'}{a}+\frac{2b'}{b}\right)\left(\frac{\phi'}{\phi}+\frac{\psi'}{\psi}\right)+1=0,
\end{eqnarray}
where a prime stands for $d/d\eta$ and $c_1$ a constant of integration.

A solution of the equation~(\ref{prime-2}) is $Z(\eta)=Z_0\sin(\eta+\eta_0)$
where $Z_0\neq0$ is an integration constant.
It should be noted that $\eta_0$ can be set equal to zero without loss of generality.
Therefore we get the following exact solutions
\begin{eqnarray}
\label{sol-1}
a(\eta)&=&a_0\left[tan\left(\frac{\eta}{2}\right)\right]^{m_1},\hspace{18mm}
b(\eta)=b_0sin\eta\left[tan\left(\frac{\eta}{2}\right)\right]^{m_2},
\\\nonumber
\\
\label{sol-3}
\phi(\eta)&=&\phi_0\left[tan\left(\frac{\eta}{2}\right)\right]^{m_3},\hspace{17mm}
\psi(\eta)=\psi_0\left[tan\left(\frac{\eta}{2}\right)\right]^{m_4},
\end{eqnarray}
where $m_i$ (with $i=1,2,3,4$) were defined as
 \begin{eqnarray}\label{c1}
m_1&\equiv&\frac{2}{3}\left(2\alpha+\beta\right), \hspace{30mm}   m_2\equiv-\frac{1}{3}\left(2\alpha+\beta\right),
\\\nonumber\\
\label{c2}
 m_3&\equiv&\beta,\hspace{45mm} m_4\equiv-\frac{2}{3}\left(\alpha+2\beta\right).
 \end{eqnarray}
In equations \eqref{sol-1}-\eqref{c2}, $a_0$, $b_0$, $\phi_0$,
$\psi_0$, $\alpha$ and $\beta\equiv\frac{c}{Z_0}$ are either integration
constants or parameters of the model.
Moreover, using~(\ref{prime-5}), we have easily shown that $a_0b_0\phi_0\psi_0=Z_0$ and
 \begin{eqnarray}\label{sol-5}
4\alpha^2&+&6\alpha\beta+\left(\frac{3\omega}{2}+5\right)\beta^2-3=0,
\end{eqnarray}
which can be rewritten as
  \begin{eqnarray}\label{omega}
\omega= -\frac{2(4\alpha^2+6\alpha\beta+5\beta^2-3)}{3\beta^2}.
  \end{eqnarray}

Furthermore, we obtained the following constraints
\begin{eqnarray}\label{Constraints-m-n}
 \sum_{i=1}^4m_i=0,\hspace{10mm}\sum_{i=1}^4m_i^2=2-\omega m_3^2,
  \end{eqnarray}
where we have used (\ref{omega}).

 Let us make some remarks about these solutions:
 \begin{itemize}
   \item
  We should note that there are only two independent parameters, so
 the third one is constrained by (\ref{sol-5}).

   \item
    Assuming $\alpha=-2\beta$, so $\psi(\eta)$ takes constant values.
Therefore, our herein solutions can be reduced to those found for a vacuum $4D$ space-time in the context of the BD cosmology.

   \item
    If $\beta$ tends to zero, then $|\omega|$ tends to infinity.
  In this particular case, the BD scalar field takes constant values. Consequently,
    the solutions (\ref{sol-1}) and (\ref{sol-3}) can be reduced to those found in an empty $5D$ space-time in GR.
 (It is worth noting that when the BD coupling parameter goes to infinity, the BD solutions may
reduce (but not always~\cite{BR93,BS97}) to their GR counterparts).
 \end{itemize}

 \section{Effective BD-KS cosmology on a four dimensional hypersurface}
\label{red-cosmology}

In this section, we give a summary of the reduced BD-KS cosmology on the hypersurface; for a more in-depth analysis, see \cite{Rasouli:2018owa}.

Applying the framework reviewed in section~\ref{MBDT},
we will obtain the components of the induced EMT and
the scalar potential. Then, we will proceed to obtain the solutions associated with the MBDT cosmology.

Using~(\ref{matt.def}) and (\ref{DO-metric}), we can easily obtain
the non-vanishing components of the induced EMT in terms of the comoving time:
\begin{eqnarray}\label{t-00}
\frac{8\pi}{\phi}T^{0[{\rm MBDT}]}_{\,\,\,0}\!&=&\!
-\frac{\ddot{\psi}}{\psi}+\frac{V(\phi)}{2\phi},\\\nonumber
\\
\label{t-ii}
\frac{8\pi}{\phi}T^{1[{\rm MBDT}]}_{\,\,\,1}\!&=&\!
-\frac{\dot{a}\dot{\psi}}{a\psi}+\frac{V(\phi)}{2\phi},
\end{eqnarray}
where replacing $a$ by $b$ in relation~(\ref{t-ii}), we get the other
component $\frac{8\pi}{\phi}T^{2[{\rm MBDT}]}_{\,\,\,2}$
(that is equal to $\frac{8\pi}{\phi}T^{3[{\rm MBDT}]}_{\,\,\,3}$).
Moreover, to obtain $V(\phi)$ we will use~(\ref{v-def}).

It is also easy to obtain the energy density $\rho$ and
pressures $P_i$ (where $i=1,2,3$) in terms of $\eta$:
\begin{eqnarray}\label{t-00-eta-general}
\rho(\eta)\equiv-T^{0[{\rm MBDT}]}_{\,\,\,0}\!&=&\!
\frac{\phi(\eta)}{8\pi b^2(\eta)}\left(\frac{\psi''}{\psi}-\frac{b'\psi'}{b\psi}\right)-\frac{V(\eta)}{16\pi},\\\nonumber
\\\nonumber
\\
\label{t-11-eta-general}
P_1(\eta)\equiv T^{1[{\rm MBDT}]}_{\,\,\,1}\!&=&\!
-\frac{\phi(\eta)}{8\pi b^2(\eta)}\frac{a'}{a}\frac{\psi'}{\psi}+\frac{V(\eta)}{16\pi},\\\nonumber
\\\nonumber
\\
\label{t-22-eta-general}
P_2(\eta)\equiv T^{2[{\rm MBDT}]}_{\,\,\,2}\!&=&\!P_3(\eta)\equiv T^{3[{\rm BD}]}_{\,\,\,3}=
-\frac{\phi(\eta)}{8\pi b^2(\eta)}\frac{b'}{b}\frac{\psi'}{\psi}+\frac{V(\eta)}{16\pi}.
\end{eqnarray}
Moreover, equation~(\ref{v-def}) for our model reduces to
\begin{equation}\label{pot-eta-all}
\frac{dV(\phi)}{d\phi}=\frac{2(1+\omega)}{b^2(\eta)}\left(\frac{\phi'}{\phi}\right)\left(\frac{\psi'}{\psi}\right).
\end{equation}

To get the energy density and pressures, we should
first obtain the induced potential.
Substituting solutions (\ref{sol-1}) and (\ref{sol-3}) into equation (\ref{v-def}), we get:
\begin{equation}\label{pot-eta}
V(\eta)=\frac{V_0}{2}\int du \left(1+ u^2\right)^4u^m,
\end{equation}
where
\begin{eqnarray}\label{m-V0}
 m&\equiv&\frac{1}{3}(4\alpha+5\beta-15),\hspace{12mm}V_0\equiv-\frac{(1+\omega)(\alpha+2\beta)\beta^2\phi_0}{12b_0^2},\\
\label{u}
u(\eta)&\equiv& \tan\left(\frac{\eta}{2}\right).
\end{eqnarray}
Equation (\ref{pot-eta}) gives
\begin{eqnarray}\label{pot-eta-2}
V(\eta)\!&=&\!V_0
u^m\sum^4_{n=0}\left[\dbinom{4}{n}
\frac{u^{2n+1}}{m+(2n+1)}\right],
 \end{eqnarray}
 where, without loss of generality, the integration constant has been set equal to zero.

Before proceeding, let us outline some particular but important cases:
\begin{itemize}
  \item
Assuming $\beta=0$, we get $\phi=\phi_0={\rm constant}$.
For this particular case, the solutions obtained in the
 previous section reduce to the corresponding KS cosmological model in GR in five dimensions.
Therefore, our herein model may reduce to
  of the KS model in the context of the IMT.

\item
 For the particular case where $\omega=-1$, we can get close resemblance between
 the scalar--tensor theories and supergravity~\cite{Faraoni.book}.

\item Assuming $\alpha=-2\beta$, $\psi(\eta)$ takes constant values, hence all the
 solutions we obtained up to now reduce to the corresponding ones
 retrieved for the standard BD theory in four dimensions.

\end{itemize}

Let us return to the general case.
To obtain the energy density and pressures, we substitute $a(\eta)$, $b(\eta)$, $\phi(\eta)$ and $\psi(\eta)$ from relations
(\ref{sol-1}) and (\ref{sol-3}) into
 (\ref{t-00-eta-general})-(\ref{t-22-eta-general}). Therefore, we obtain the components of
the induced EMT in terms of $\eta$
\begin{eqnarray}\label{t-00-eta}
\rho(\eta)\!\!&=&\!\!T_0\left\{
\Big[(\beta+2)+(\beta-2) u^2\Big]\left(1+ u^2\right)^3
+(1+\omega)\beta^2\sum^4_{n=0}\dbinom{4}{n}
\frac{u^{2n+1}}{m+(2n+1)}\right\}u^{m},
\\\nonumber
\\\nonumber
\\
\label{t-11}
P_1(\eta)\!\!&=&\!\!T_0\left\{\left[\frac{2}{3}(2\alpha+\beta)\right]u\left(1+ u^2\right)^4
-(1+\omega)\beta^2\sum^4_{n=0}\dbinom{4}{n}
\frac{u^{2n+1}}{m+(2n+1)}\right\}u^{m},
\\\nonumber
\\\nonumber
\\
\label{t-22}
P_2(\eta)&=&\frac{T_0}{3}\left[(3-2\alpha-\beta)-(3+2\alpha+\beta) u^2\right]u^{m+1}
\left(1+ u^2\right)^3\cr
\cr
&-&T_0\,(1+\omega)\beta^2u^{m}\sum^4_{n=0}\dbinom{4}{n}
\frac{u^{2n+1}}{m+(2n+1)},
\end{eqnarray}
where
\begin{eqnarray}\label{T0}
T_0\equiv\frac{\phi_0(\alpha+2\beta)}{192\pi b_0^2},
 \end{eqnarray}
and we have used (\ref{pot-eta-2}).

We should note that as $T^{1[{\rm MBDT}]}_{\,\,\,1}\neq T^{2[{\rm MBDT}]}_{\,\,\,2}$, therefore,
the induced matter is not a perfect fluid.
In the noncompactified KK frameworks described
above, it has been shown that the induced EMT obeys the conservation law
that is for our herein model is given by
\begin{eqnarray}
\label{EMT-Cons-1}
\dot{\rho}+\sum^3_{i=1}\left(\rho+P_i\right)H_i=0,
\end{eqnarray}
where $H_1=\dot{a}/a$ and $H_2=H_3=\dot{b}/b$  denote the directional Hubble parameters.
Using \eqref{sol-1} and (\ref{sol-3}), equation (\ref{EMT-Cons-1}), in terms of the new time coordinate, is rewritten as
\begin{eqnarray}
\label{EMT-Cons-2}
u\rho'(\eta)+\frac{1}{3}\left(2\alpha +\beta\right)\left(1+\zeta u^2\right)\Big[P_1(\eta)-P_2(\eta)\Big]
+\left(1-\zeta u^2\right)\Big[\rho(\eta)+P_2(\eta)\Big]
=0,
\end{eqnarray}
where $u(\eta)$ is given by (\ref{u}). Substituting the energy density and the
pressures from (\ref{t-00-eta})-(\ref{t-22}) into (\ref{EMT-Cons-2}) and then
employing equation (\ref{sol-5}), it has been shown that \eqref{EMT-Cons-2} is satisfied for
for our herein both BD-KS model \cite{Rasouli:2011rv}.

In what follows, let us study the properties of some
  physical quantities such as the average
  Hubble parameter ${\rm H}$, the deceleration
parameter $q$, the spatial volume ${\rm V}_s$, mean anisotropy
parameter ${\rm A}_h$, and the expansions for scalar
expansion $\theta$ and the shear scalar $\sigma^2$:
\begin{eqnarray}\nonumber
V_s\!\!&=&\!\!A^3(t)=a(t)b^2(t), \hspace{5mm}\theta=3H=\left(\frac{\dot{a}}{a}+\frac{2\dot{b}}{b}\right),\\\nonumber
A_h\!\!&=&\!\!\frac{1}{3}\sum^3_{i=1}\left(\frac{\Delta H_i}{H}\right)^2, \hspace{5mm} {\rm where}\hspace{5mm}\Delta H_i=H_i-H,\\\nonumber
q\!\!&=&\!\!\frac{d}{dt}\left(\frac{1}{H}\right)-1=-\frac{A\ddot{A}}{\dot{A}^2},\\
\label{phys.quant}
\sigma^2\!\!&=&\!\!\frac{1}{2}\sigma_{ij}\sigma^{ij}=\frac{1}{3}\left(H_1^2+H_2^2-2H_1H_2\right),
\end{eqnarray}
where $i,j=1,2,3$ and ${\rm A(t)}$ denotes the mean scale factor of the universe.
By substituting $a(\eta)$, $b(\eta)$, $\phi(\eta)$ and $\psi(\eta)$ from relations
(\ref{sol-1}) and (\ref{sol-3}) into (\ref{phys.quant}), we get
\begin{eqnarray}\nonumber
V_s(\eta)\!\!&=&\!\!A^3(\eta)=a_0b_0^2\left(\frac{2u}{1+ u^2}\right)^2,\\\nonumber
\\\nonumber
\theta(\eta)\!\!&=&\!\!3H(\eta)=\left(\frac{1-u^4}{2b_0 u^2}\right)u^{\frac{1}{3}(2\alpha+\beta)},\\\nonumber
\\\nonumber
A_h(\eta)\!\!&=&\!\!(2\alpha+\beta)\left(\frac{1+ u^2}{1- u^2}\right)
\left[\left(\frac{2\alpha+\beta}{2}\right)\left(\frac{1+ u^2}{1- u^2}\right)-1\right]+\frac{1}{2},\\\nonumber
\\\nonumber
q(\eta)\!\!&=&\!\!\frac{(4+2\alpha+\beta)u^4+4(1+ u^2)-(2\alpha+\beta)}{2(1-u^2)^2},\\\nonumber
\\\label{phys.quant-2}
\sigma^2(\eta)\!\!&=&\!\!\frac{1}{3b_0^2}\left(\frac{1+ u^2}{2u}\right)^4
\left[(2\alpha+\beta)-\left(\frac{1- u^2}{1+ u^2}\right)\right]^2u^{\frac{2}{3}(2\alpha+\beta)}.
\end{eqnarray}
It is worthy to note that when $\omega$ goes to infinity, our herein BD-KS solutions may reduce to the corresponding
 ones in IMT. For instance, as seen from relation (\ref{omega}), when $\beta$ tends
 to zero, then $\vert\omega\vert$ goes to infinity.
 Therefore, for this particular case, we get $\phi=\phi_0={\rm constant}$ and
\begin{equation}\label{101}
\alpha=\pm\frac{\sqrt{3}}{2},
 \end{equation}
where we have used~(\ref{sol-5}).
In this case, setting $\beta=0$ and substituting $\alpha$
from \eqref{101} into (\ref{sol-1}) and (\ref{sol-3}), we obtain
 \begin{eqnarray}
\label{IMT-sol-2}
a(\eta)=a_0u^{\pm\frac{2\sqrt{3}}{3}},\hspace{5mm}
b(\eta)=b_0sin\eta \,\,u^{\mp\frac{\sqrt{3}}{3}},\hspace{5mm}
\psi(\eta)=\psi_0u^{\mp\frac{\sqrt{3}}{3}},
\end{eqnarray}
where $u=u(\eta)$ was given by (\ref{u}).
Obviously, equations (\ref{prime-2})-(\ref{prime-5}) are satisfied by substituting
these solutions, where assuming $c_1=0$, it is seen that (\ref{prime-1}) yields an identity, $0=0$.
Moreover, for this particular case, the induced scalar potential
vanishes, see (\ref{pot-eta}). Therefore, employing (\ref{t-00-eta})-(\ref{t-22}), we obtain
\begin{eqnarray}\label{t-00-IMT}
\rho(\eta)\!&=&\!2 T_0\left(1-\zeta u^2\right)\left(1+\zeta u^2\right)^3
u^{m+1},
\\\nonumber
\\\nonumber
\\
\label{t-11-IMT}
P_1(\eta)\!&=&\!
\pm\frac{2\sqrt{3}}{3}T_0\left(1+\zeta u^2\right)^4u^{m+1},\\\nonumber
\\\nonumber
\\
\label{t-22-IMT}
P_2(\eta)\!&=&\!
\frac{T_0}{3}\left[(3\mp\sqrt{3})-3(3\pm\sqrt{3})\zeta u^2\right]
\left(1+\zeta u^2\right)^3u^{m+1},
\end{eqnarray}
where
\begin{eqnarray}\label{T0}
T_0\equiv\frac{\pm \sqrt{3}\phi_0}{384\pi b_0^2},
\hspace{10mm}m=\frac{1}{3}(\pm 2\sqrt{3}-15).
\end{eqnarray}

Finally, we should note that for our herein anisotropic model, retrieving the
analytical solutions in terms of the cosmic time is a complicated procedure.

\section{Conclusions and discussions}
\label{Conclusions}

The development of KK theory has a lengthy and significant history.
The cylinder condition is one of the fundamental flaws in Kaluza's primary mechanism.
Moreover, the compactness condition of the
Klein's version has produced certain physical problems \cite{MWL98}.
Many attempts have been made to address the problems
with the original KK theory by modifying their underlying assumptions.
One of the more intriguing contemporary versions
of the original KK theory is the IMT, which avoids recurring
issues by using a completely general form for the underlying
metric. We have provided the main formalism
of the IMT in Section \ref{IMT} of this review.
The interpretation of the field equations associated with the IMT and
their applications in cosmology and astrophysics have been
extensively presented in the literature, see
for instance,~\cite{OW97,wesson1999space,RRT95,W08,I09,RB09,BCF13,WO13,M15}.

The MBDT \cite{RFM14} and MSBT \cite{RM18,RPSM20,RJM22} are two significant
instances of recently established generalized IMT versions.
The underlying theories for these updated frameworks replace GR with the BD and SB theories.
Higher-dimensional matter fields were also taken into account when
developing these modified theories, resulting in more generalized frameworks.
In the MBDT and MSBT contexts, not only matter fields but also an induced
scalar potential emerge as a result of the existence and curvature of the
extra dimension, which distinguishes these theories from their standard counterparts.
 To be more specific, let us concentrate on the BD theory.
In the generalized versions of the standard BD theory, several
ad hoc assumptions (such as varying $omega$ \cite{BP01} and manually adding
potential to the action \cite{SS03,MM23}) have been made in order to depict evolution
of our 4D universe, specifically for attaining
accelerating scale factor \cite{Rasouli:2014dba,Rasouli:2016syh,Rasouli:2018lny}.
However, in the MBDT, the fundamental induced potential as
well as induced matter play the necessary role \cite{RFM14}.
Therefore, for the cosmological models in the contexts of the MBDT and MSBT we do not need any phenomenological assumptions.

Aside from the FLRW model \cite{Rasouli:2016ngl}, anisotropic cosmological models have
been studied in the context of GR or the alternative theories to
GR (in 4D as well as higher-dimensional contexts) \cite{TB21,P22,LGLMM21,PSB18,AKOV20,ADK22},
and noncompactified KK theories \cite{PW08,Rasouli:2011rv,Rasouli:2014sda,Rasouli:2018owa}.
In this paper, we presented the anisotropic BD-KS model as an application of the MBDT.

Last but not least, using the MSBT and MBDT,  due to their
enormous potential, in future studies to investigate a number of open problems will be beneficial.

\section*{Acknowledgments}
I would like to thank the organizers of \textit{`Sixth International Conference on the Nature and Ontology of Spacetime'}.
I also acknowledges the FCT grants UID-B-MAT/00212/2020 and UID-P-MAT/00212/2020
at CMA-UBI plus the COST Action CA18108 (Quantum gravity phenomenology in the multi--messenger approach).

\bibliographystyle{utphys}

\end{document}